\documentclass[epj,final]{svjour}

\usepackage{amsmath}
\usepackage{amssymb}
\usepackage{graphicx}

\DeclareGraphicsExtensions{.eps}
\graphicspath{{./figures/}}
\newlength{\figwidth}

\begin{document}
\title{Conductance of nano-systems with interactions coupled via 
conduction electrons: Effect of indirect exchange interactions}
\titlerunning{Indirect exchange interaction}

\author{%
Yoichi Asada\inst{1} 
\and 
Axel Freyn\inst{2} 
\and 
Jean-Louis Pichard\inst{2}
}
\authorrunning{Y.\ Asada \textit{et al.}}

\institute{%
Department of Physics, Tokyo Institute of Technology, 2-12-1 Ookayama, 
Tokyo 152-8551, Japan 
\and 
Service de Physique de l'Etat Condens{\'e} (CNRS URA 2464),
DSM/DRECAM/SPEC, \\ 
CEA Saclay, 91191 Gif sur Yvette Cedex, France
}

\date{\today}

\abstract{%
 A nano-system in which electrons interact and in contact with 
Fermi leads gives rise to an effective one-body scattering which 
depends on the presence of other scatterers in the attached leads. 
This non local effect is a pure many-body effect that one neglects 
when one takes non interacting models for describing quantum transport. 
This enhances the non-local character of the quantum conductance by 
exchange interactions of a type similar to the RKKY-interaction between 
local magnetic moments. A theoretical study of this effect is given 
assuming the Hartree-Fock approximation for spinless fermions of Fermi 
momentum $k_{\mathrm{F}}$ in an infinite chain embedding two scatterers 
separated by a segment of length $L_c$. The fermions interact only inside 
the two scatterers. The dependence of one scatterer onto the other exhibits 
oscillations of period $\pi/k_{\mathrm{F}}$ which decay as $1/L_c$ and 
which are suppressed when $L_c$ exceeds the thermal length 
$L_\mathrm{T}$. The analytical results given by the Hartree-Fock 
approximation are compared with exact numerical results obtained with 
the embedding method and the DMRG algorithm. 
}

\maketitle

%%%%%%%%%%%%%%%%% INTRODUCTION %%%%%%%%%%%%%%%%%%%%%%

\section{Introduction} 

  The coupling of nano-objects via conduction electrons was discovered 
long ago, in the case of spins of magnetic ions, or of nuclei, which 
results indirectly from the interaction of such spins with those of 
conduction electrons in metals. After tracing out the degrees of freedom 
of the conduction electrons, one gets an effective spin Hamiltonian 
characterized by an oscillatory long range interaction, 
the RKKY-interaction \cite{RK,Y,VV,BF}, which plays a crucial role in 
understanding the large variety of possible ordered spin structures in 
magnetic crystals. If $\mbox{\boldmath{$I$}}_i$ is a local magnetic moment 
in a metal, the conduction electrons give rise to an interaction energy 
between these moments, which can be described by an Hamiltonian of the form: 
\begin{equation}
H_{\mathrm {RKKY}}=\sum_i \sum_{j<i} J_{ij} \mbox{\boldmath{$I$}}_i
\cdot \mbox{\boldmath{$I$}}_j.
\end{equation}
The coupling term between two moments separated by a distance 
$R_{ij}$ behaves as 
\begin{equation}
J_{ij} \propto \frac{2k_{\mathrm{F}} R_{ij} \cos (2k_{\mathrm{F}} R_{ij})- \sin (2k_{\mathrm{F}} R_{ij})}
{R^4_{ij}}
\end{equation}
in $d=3$ dimensions, $k_{\mathrm{F}}$ being the Fermi momentum of the conduction 
electrons. In $d=1$ dimension, this gives a long range $1/R$ interaction 
with oscillations of periodicity $\pi/k_{\mathrm{F}}$. 

We show in this work that a similar phenomenon characterizes also 
the quantum conductance of nano-systems in which electrons 
interact, coupled via metallic wires. This phenomenon is very general 
and does not require to include the spin degrees of freedom. Combining  
Landauer's formulation \cite{Landauer} of quantum transport and the 
Hartree-Fock approximation, as reviewed in Ref. \cite{Datta,Asada},
we will show that the scattering matrix of an interacting system depends on 
what is embedded at a distance $R$ of the interacting system in the attached 
leads, this dependence decaying as $1/R^d$ with oscillations of periodicity  
$\pi/k_{\mathrm{F}}$. This non local character of the quantum conductance 
is another example of the effect of indirect exchange interactions between 
interacting nano-systems via conduction electrons, as the RKKY-interaction 
between local magnetic moments. 

   To study this phenomenon, we take an infinite chain where spinless 
fermions do not interact outside two identical regions where the scattering 
is a pure many body effect due to Coulomb repulsions. The chain is described 
by a one dimensional tight binding model with nearest neighbor hopping 
$t_h=1$. A nearest neighbor repulsion of strength $U$ acts only between 
two consecutive sites, in two regions which are connected by $L_{\mathrm c}$ 
sites where the particles do not interact. This model is simple enough to 
be analytically solved at a mean field level, using the Hartree-Fock (HF) 
approximation. To simplify the calculations, we add positive compensating 
potentials which exactly cancels the Hartree terms of the HF equation. But 
the exchange term modifies the hopping term coupling the two internal sites 
of each scatterers, in such a way that it takes a value $v$ instead of $t_h$. 
In the HF approximation, $v$ is given by a self-consistent equation. Having 
$v$, it is straightforward to obtain the scattering matrix $S_1(v)$ of one 
scatterer at the Fermi energy of the infinite non interacting chain. Using 
the combination law of the one-body $S$-matrices in series, one can get the 
scatteringmatrix $S_2(v)$, and hence the dimensionless conductance 
$g_2^{k_{\mathrm{F}}}$ (in units of $e^2/h$) of these two many-body 
scatterers in series, in the HF approximation. 

 For such a system where two identical many-body scatterers are 
coupled by a perfect wire of $L_{\mathrm c}$ sites where the electrons 
do not interact, our main result is to show that the value of $v$ 
characterizing one scatterer differs from its value when there is no 
second scatterer by an oscillatory term of period $\pi/k_{\mathrm{F}}$ which 
decays as $1/L{\mathrm c}$. In certain limits, $v$ can be given in 
a simple form as a function of $U,L_{\mathrm c}$ and $k_{\mathrm{F}}$. For instance, 
for a half-filled chain ($k_{\mathrm{F}}=\pi/2$) and weak interactions, the effective 
hopping term $v$ reads 
\begin{equation}
v = 1 + \frac{2}{2-\pi+2\pi/U} +\frac{(-1)^{L_{\mathrm c}}}
{L_{\mathrm c}} \left( \frac{1}{2-\pi+2\pi/U} \right)^2 + \ldots 
\label{approxbis}
\end{equation}
In this limit, this expression shows how the effective hopping 
term $v$ characterizing a single scatterer decays as $1/L_c$ towards 
its value when it is not in series with another one, with the even-odd 
oscillations characteristic of a half-filled chain. Using 
Eq. \eqref{approxbis}, it is straightforward to show how the scattering
matrix 
$S_1(v)$ of a single scatterer is modified by the presence of a second 
scatterer.  We underline that this non local effect is a pure many-body 
effect that one neglects when one takes non interacting models for 
describing quantum transport. This non local effect was first numerically 
discovered in a previous work \cite{MWP}, using the embedding method 
\cite{Molina1,Molina2,Molina3,Gogolin,Mila,Sushkov,Meden,Rejec} and the DMRG 
algorithm \cite{DMRG1,DMRG2} valid for one dimensional fermions. In this 
work, we give a simple theory of this effect based on the HF approximation, 
which turns out to qualitatively describe this non local effect for all 
values of $U$, including its suppression in the limit when $U \rightarrow 
\infty$. The HF approximation becomes quantitatively accurate for small 
strengths $U$ of the interaction. Moreover, we will also show that this non 
local effect vanishes when the length $L_{\mathrm c}$ of the coupling wire 
exceeds the thermal length $L_{\mathrm T}$ characterizing free fermions in 
one dimension.

The paper is organized as follows: 

In section \ref{section2}, we consider a single scatterer with a nearest 
neighbor repulsion of strength $U$ embedded in an infinite chain. In the 
first sub-section, the HF equation of this simple model is written, 
leading us to study a chain where the hopping term between the two central 
sites is equal to $v$ instead of $t_h=1$. This one body model is solved in 
the second sub-section, allowing us to obtain the implicit 
equation giving $v$ in the HF approximation. The system being symmetric 
upon reflection, we use scattering phase shifts and Friedel sum rule for 
this purpose. In the third sub-section, the conductance of this single 
scatterer is studied as a function of the strength $U$ of the nearest 
neighbor repulsion, the HF behavior being compared to the exact results 
given by the embedding method and the DMRG algorithm. In the fourth 
sub-section, the Friedel oscillations of the particle density around the 
scatterer are described. In a last sub-section, a correlation function 
inside the attached leads is calculated at a distance $p$ from the 
scatterer, which will be useful for describing the case of two scatterers 
in series. This function is shown to decay as $1/p$ with oscillations of 
periodicity $\pi/k_{\mathrm{F}}$ towards an asymptotic value characterizing 
the chain without scatterer. 

 In section \ref{section3}, we study the conductance of two scatterers 
in series, coupled by a scattering free wire of $L_c$ sites. 
In the first sub-section, simple analytical expressions are given in the 
limit where the strength $U$ of the interaction acting inside the two 
scatterers remains small. Notably, we show that the effective hopping 
term $v$ characterizing each scatterer differs from the value obtained 
in the section \ref{section2} by a correction which decays as 
$1/L_c$ with oscillations of periodicity $\pi/k_{\mathrm{F}}$. In the 
second sub-section, the Hartree-Fock equation is solved exactly for  
arbitrary values of $U$, allowing us to show that the weak $U$-expansion 
assumed in the first sub-section remains valid till $U \approx t_h$. 
The effective hopping term $v$ characterizing a single scatterer being 
modified when it is in series with another, the implication of this non 
local effect upon the conductance is illustrated in the third sub-section, 
the HF curves describing the conductance oscillations of the two 
scatterers in series being similar to the curves numerically obtained in 
Ref. \cite{MWP}. 

 In section \ref{section4}, the results for two scatterers in series given 
by the HF theory are compared to the exact numerical values given in 
Ref. \cite{MWP}. The HF theory turns out to give a good qualitative 
description of the non local effect, which becomes quantitatively accurate 
when $U < t_h$ for the very small scattering regions which we have considered.
 
In section \ref{section5}, we show that this non local effect is suppressed 
at a temperature $T$, when the length $L_c$ of the coupling wire 
exceeds the thermal length $L_{\mathrm T} \propto v_{\mathrm F}/T$ of free 
fermions in one dimension, $v_{\mathrm F}$ being the Fermi velocity. 

 We conclude in section \ref{section6} by a  summary of the main results, 
underlining their relevance for a theory of the non local character 
of quantum transport measurements, and suggesting straightforward 
extensions of this theory outside one dimension.   

\section{Transmission through a single many-body scatterer}
\label{section2}

\subsection{Microscopic model and exchange energy}

To study the indirect exchange interaction via conduction electrons 
between nano-systems in which the electrons interact, we begin to study 
the simplest many-body scatterer, taking a tight-binding model of 
spinless fermions which do not interact, unless they occupy the 
two central sites $0$ and $1$ of an infinite chain, which costs a 
nearest neighbor repulsion energy $U$. Assuming the Hartree-Fock 
approximation, this leads us to study an analytically 
solvable one body model where the hopping term between the two central 
sites is modified  by the interaction, modification which has to be 
calculated self-consistently. The many-body scattering system is described 
by an Hamiltonian $H=H_{kin}+H_{int}$. The kinetic and interaction terms 
respectively read
\begin{equation}
\begin{aligned}
H_{kin} = & -\sum_{p=-\infty}^{\infty} t_\mathrm{h} 
(c^\dagger_p c^{\phantom{\dagger}}_{p-1} + c^\dagger_{p-1}
c^{\phantom{\dagger}}_p)\\
H_{int} = & \ U \left[n_1-V_+\right]\left[n_{0}-V_+\right] \, .
\end{aligned}
\label{hamiltonian}
\end{equation}
The hopping amplitude $t_\mathrm{h}=1$ between nearest neighbor sites 
sets the energy scale, $c^{\phantom{\dagger}}_p$ ($c^\dagger_p$) is 
the annihilation (creation) operator at site $p$, and 
$n_p = c^\dagger_p c^{\phantom{\dagger}}_p$. The potential 
$V_+$ is due to a positive background charge which exactly 
cancels the Hartree term of the HF equation. The conduction band 
corresponds to energies $-2 <E=-2\cos k <2$ ($k$ real). $H$ is invariant 
under reflections ($p-1/2 \rightarrow -p+1/2$) and exhibits particle-hole 
symmetry if the chain is half filled. In this case, $V_+=1/2$, the Fermi 
momentum $k_{\mathrm{F}}=\pi/2$ and one has a uniform density without 
Friedel oscillations around the central region where the fermions interact.  
 
   In the HF approximation, one assumes a variational ground state 
which is a Slater determinant of one-body wave-functions 
$\psi_{\alpha}(p)$ of energies $E_{\alpha}<E_F$. 
Since in our model the negative charge inside the scatterer is exactly 
compensated by a positive background charge, the Hartree term is cancelled  
and we have just to take into account the exchange term in the Hartree-Fock 
equation \cite{Ashcroft-Mermin,Fetter-Walecka} giving the $\psi_{\alpha}(p)$:
\begin{equation}
-\psi_{\alpha}(p+1)-\psi_{\alpha}(p-1)
-\sum_{p'} t_{p,p'}^{\rm HF} \psi_{\alpha}(p')
= E_{\alpha} \psi_{\alpha}(p).
\end{equation}
The exchange term   
\begin{equation}
t_{p,p'}^{\rm HF}= \sum_{E_{\beta}<E_F} U_{p,p'} \psi_{\beta}^*(p')
\psi_{\beta}(p)
\end{equation}
with the taken nearest neighbor repulsion 
\begin{equation}
U_{p,p'}=U \left \{ \delta_{p,0}\delta_{p',1}+
\delta_{p,1}\delta_{p',0} \right \},  
\end{equation} 
is very simple. It is zero, excepted between the two central sites 
where the fermions interact, where it yields an increase of the 
strength of the hopping term coupling the two central sites by an amount 
\begin{equation}
t_{0,1}^{\rm HF}(U) = U \sum_{E_{\alpha}<E_F} 
\psi^*_{\alpha}(1) \psi_{\alpha}(0)
= U \left\langle c^\dagger_1 c^{\phantom{\dagger}}_{0} 
\right \rangle. 
\end{equation}

\begin{figure}
\centerline{\includegraphics[width=\linewidth]{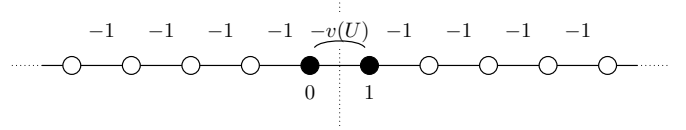}}
\caption{Effective one body model obtained assuming the Hartree-Fock 
approximation for a single many body scatterer (Hamiltonian 
\eqref{hamiltonian}). The hopping term (indicated above) $v(U)$ 
between the sites at $p=0$ and $p=1$ (indicated below)
depends on $U$ and $k_{\mathrm{F}}$.}
\label{Fig1}
\end{figure}

The Hartree-Fock equation describes a tight-binding model which is 
represented in Fig.\ref{Fig1}, where the hopping term between 
the two central sites is no longer equal to $t_h=1$, but takes 
an interaction dependent value $v$, which is given by an implicit 
equation: 
\begin{equation}
v = t_h + t_{0,1}^{\rm HF}(v)= 1 + t_{0,1}^{\rm HF}(v)
\label{def-v}
\end{equation}
for $t_h=1$.

\subsection{Scattering phase shifts and density of states}

 The effective one-body model being symmetric upon the reflection 
$(p-1/2\rightarrow -p+1/2)$, its Hamiltonian has even and odd standing-wave 
solutions $\psi_k^0 (p)$ and  $\psi_k^1 (p)$, which can be written 
\cite{Lipkin} inside the conduction band as: 
\begin{equation}
\begin{aligned}
\psi_k^0 (p) = & \sqrt{\frac{2}{L}} \cos 
\left(k(p-\frac{1}{2})-\delta_0(k)\right) \\
\psi_k^1 (p) = & \sqrt{\frac{2}{L}} \sin
\left(k(p-\frac{1}{2})-\delta_1(k)\right)  ,
\end{aligned}
\end{equation}
at the left side of the scatterer ($p\leq 0$) and
\begin{equation} 
\begin{aligned}
\psi_k^0 (p) = & \sqrt{\frac{2}{L}} \cos 
\left(k(p-\frac{1}{2})+\delta_0(k)\right) \\
\psi_k^1 (p) = & \sqrt{\frac{2}{L}} \sin
\left(k(p-\frac{1}{2})+\delta_1(k)\right)  ,
\end{aligned}
\end{equation}
at its right side ($p\geq1$). The normalization factor 
$\sqrt{\frac{2}{L}}$ corresponds to a chain of length 
$L \rightarrow \infty$. The scattering when $v \neq 1$ 
gives rise to two phase shifts $\delta_0(k)$ and $\delta_1(k)$. 
Writing the Schr\"odinger equation  inside the scatterer (sites 
$0$ and $1$) for the even and odd solutions, one gets:   
\begin{equation}
\begin{aligned}
-( 2\cos k) \psi_k^{0,1} (0) = & - \psi_{k}^{0,1} (-1) -v \psi_k^{0,1} (1) \\
-( 2\cos k) \psi_k^{0,1} (1) = & - v \psi_{k}^{0,1} (0) - \psi_k ^{0,1}(2),  
\end{aligned}
\end{equation}
which yields the following expressions for the even and odd phase shifts:
\begin{equation}
\tan \delta_0 (k) = \frac{v-1}{v+1} \cot \left(\frac{k}{2}\right) 
\end{equation}
and 
\begin{equation}
\tan \delta_1 (k) = \frac{1-v}{v+1} \tan \left(\frac{k}{2}\right). 
\end{equation}

 In addition, a value of $v > 1$ gives rise to two bound states 
located in the central region with energies outside the conduction band. 
The first one of energy  
\begin{equation}
E_{\rm bs1}=-(v+v^{-1}) 
\end{equation}
has an even wave-function given by 
\begin{equation}
\psi_{\rm bs1}(p)=\left(\sqrt{\frac{v-v^{-1}}{2}}\right) v^{-|p-1/2|}, 
\end{equation}
while the second of energy 
\begin{equation}
E_{\rm bs2}=v+v^{-1}
\end{equation}
has an odd wave-function given by 
\begin{equation}
\psi_{\rm bs2}(p)=\left(\sqrt{\frac{v-v^{-1}}{2}}\right) 
 (-1)^{p}  v^{-|p-1/2|}.
\end{equation}

  When a scatterer is introduced in the chain ($v \neq 1$), this yields 
a correction $\delta \rho (E)$ to the density of states $\rho(E)$, 
which is given \cite{Hewson} for the even and odd components inside the 
conduction band by: 
\begin{equation}
\delta \rho_{0,1}(E)= 
\frac{1}{\pi} \frac{\partial \delta_{0,1}(E)}{\partial E}. 
\end{equation}
$\delta \rho (E)$ satisfies \cite{Hewson} Friedel sum rule: 
\begin{equation}
\frac{\delta_{0}(E_F)+\delta_{1}(E_F)}{\pi}=
\int_{-\infty}^{E_F} \delta \rho (E) dE = \nu L_S
\end{equation}
for a scatterer of length $L_S$ embedded in a chain with a uniform 
filling factor $\nu$. This implies that 
\begin{equation}
\frac{\int_{-\infty}^{E_F} \delta \rho (E) dE }
{\int_{-\infty}^{E_F} \rho (E,v=1) dE} = \frac{L_S}{L} \rightarrow 0 
\end{equation}
when $L \rightarrow \infty$: The change $\delta \rho (k)$ of the 
density of real momenta $k$ (inside the conduction band) vanishes 
in the limit of infinite lead length. When $L \rightarrow \infty$, 
the only change in the density $\rho(k)$ due to the scatterer comes 
from the bound states $\psi_{\rm bs}$ of imaginary momenta $K=\mathrm{i}k$ which are 
occupied outside the conduction band at zero temperature. For an 
arbitrary function $F(k)$ at zero temperature, this gives the relation:
\begin{equation}
\sum_{k<k_{\mathrm{F}}} F(k)= \frac{L}{2\pi} \int_{0}^{k_{\mathrm{F}}}  F(k) dk + 
\sum_{\rm bs} F(K_{\rm bs}),     
\label{sum-int}
\end{equation} 
the last term being a sum over the occupied bound states.

\subsection{$\langle c_{1}^{\dagger}c_0 \rangle$ and 
conductance $g_1^{k_{\mathrm{F}}}(v)$}

 To obtain the implicit equation giving $v$, we need to calculate 
$\langle c_1^{\dagger}c_0 \rangle$, which is the sum of 
contributions $A_{\rm cb}^{1,0}$ due to the conduction band and 
$A_{\rm bs}^{1,0}$ due to the occupied bound states. Assuming that $E_F<2$, 
only the even bound state is occupied and:
\begin{equation}
\begin{aligned}
\left\langle c_1^{\dagger}c_0 \right\rangle= & 
A_{\rm cb}^{1,0} + A^{1,0}_{\rm bs1} \\
A^{1,0}_{\rm cb} = & \frac{L}{2\pi} \int_{0}^{k_{\mathrm{F}}} dk  
\left\{ \sum_{i=0}^1 \psi_k^{i*} (1) \psi_k^i (0) \right\} \\
= & \int_{0}^{k_{\mathrm{F}}} \frac{dk}{\pi} \left\{ \frac{4v \cos k \sin^2 k}
{1+v^4-2v^2 \cos (2k)} \right\} \\
= & \frac{v^{-2}-1}{2\pi} \arctan \left(\frac{2v \sin k_{\mathrm{F}}}{v^2-1} \right)+ 
\frac{\sin k_{\mathrm{F}}}{\pi v},  \\
A_{\rm bs1}^{1,0}= & \psi_{\rm bs1}^* (1) \psi_{\rm bs1} (0) = 
\frac{1-v^{-2}}{2}
\label{inside}
\end{aligned}
\end{equation}

 Using this, one can calculate $v$ as a function of the interaction 
strength $U$ and of the Fermi momentum $k_{\mathrm{F}}$ by solving the 
implicit equation 
\begin{equation}
v=1+ U \langle c_{1}^{\dagger}c_{0}(v) \rangle.
\label{SCHF-single}
\end{equation}
 Once the change of the effective hopping term $v$
between the two central sites 
is obtained, it is straightforward to determine the transmission 
coefficient $t_1(v)$. At zero 
temperature, the Landauer conductance $g_1^{k_{\mathrm{F}}}(v)$ (in units of 
$e^2/h$) of the central region where the electrons interact is given 
by this effective one-body transmission coefficient 
$|t_1^{k_{\mathrm{F}}}(v)|^2$. Using the Landauer formula, one gets the 
transmission coefficient and the dimensionless conductance
\begin{equation}
\begin{aligned}
t_1^{k_{\mathrm{F}}} (v) =& 
\frac{v(\mathrm{e}^{2\mathrm{i}k_{\mathrm{F}}}-1)}
{v^2-\mathrm{e}^{-2\mathrm{i}k_{\mathrm{F}}}}\\
g_1^{k_{\mathrm{F}}} (v) =& |t_1^{k_{\mathrm{F}}}(v)|^2 
= \frac{4v^2\sin^2 k_{\mathrm{F}}}{v^4-2v^2\cos(2k_{\mathrm{F}})+1}. 
\label{g1}
\end{aligned}
\end{equation}

\begin{figure}
\centerline{\includegraphics[width=\linewidth]{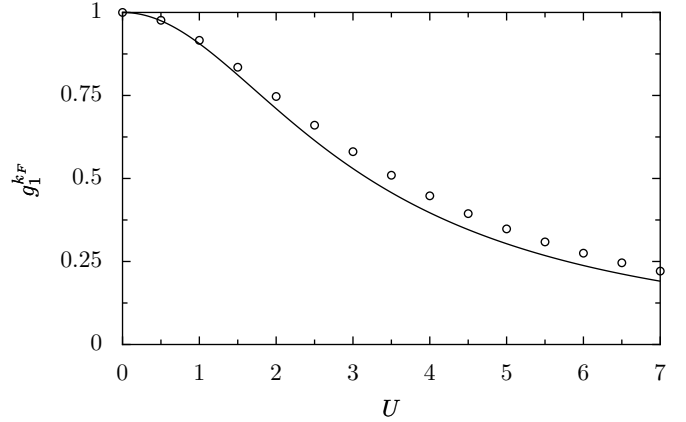}}
\caption{Conductance $g_1^{k_{\mathrm{F}}}$ of a single scatterer as a 
function of the interaction strength $U$ for $k_{\mathrm{F}}=\pi/2$. 
The solid line gives the HF behavior (Eq. \eqref{g1}). 
The circles are the exact results obtained with embedding 
method and the DMRG algorithm.}
\label{Fig1bis}
\end{figure}

The behavior of $g_1^{k_{\mathrm{F}}}$ as a function of the interaction 
strength $U$, obtained assuming the Hartree-Fock approximation 
(Eq. \eqref{g1}) is shown in Fig. \ref{Fig1bis} for $k_{\mathrm{F}}=\pi/2$. 
An accurate value for $g_1^{k_{\mathrm{F}}}$ can be obtained using the 
embedding method and the DMRG algorithm, as introduced in Refs. 
\cite{Molina1,Molina2}. Using this exact method, we have also calculated 
$g_1^{k_{\mathrm{F}}}$ at different interaction strengths $U$. The data 
are presented in Fig. \ref{Fig1bis}, showing that the HF approximation 
is a good approximation for a very short interacting region of moderate 
interaction strength $U$.

\subsection{Friedel oscillations of the density outside half-filling}

\begin{figure}
\centerline{\includegraphics[width=\linewidth]{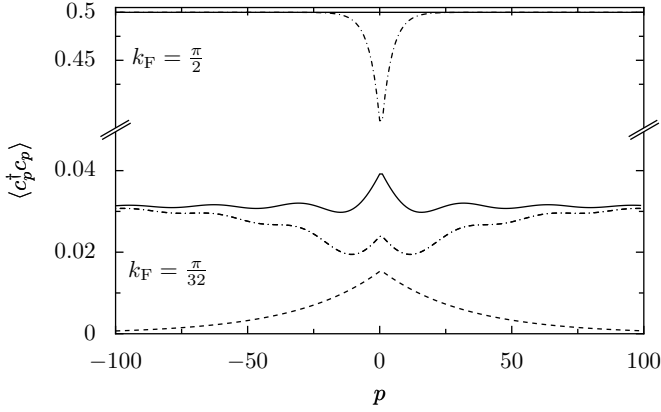}}
\caption{Particle density $n_p=\langle c_{p}^{\dagger}c_p \rangle$ 
at sites $p$ around a scatterer with an interaction strength $U=0.4$. 
For $k_{\mathrm F}=\pi/32$ (lower part of the figure): Contributions 
of the bound state (dashed line) and of the conduction band (dotted-dashed 
line) to $n_p$ and Friedel oscillations of $n_p$ around the average 
filling $1/32$ (solid line). For $k_{\mathrm F}=\pi/2$ (upper part of 
the figure): Contribution of the conduction band (dotted-dashed 
line) and total uniform density $n_p=1/2$.}
\label{density}
\end{figure}

 For a half-filled chain, the system has particle-hole symmetry, and the 
density $n_p=\langle c_{p}^{\dagger}c_p \rangle$ must be equal to $1/2$ 
at each site $p$. This is due to two opposite effects which compensate 
each other: a decrease of the contribution of the conduction band to 
the density that one can expect when there is a local repulsion acting 
inside the scatterer, and the contribution of the interaction-induced 
bound state to the density. These contributions are plotted separately 
in Fig. \ref{density}. For half-filling (upper part of Fig. \ref{density}), 
one gets an exact compensation of the two opposite effects for having a 
uniform density, and particle-hole symmetry is satisfied. Outside 
half-filling (lower part of Fig. \ref{density}),
one gets large Friedel oscillations of the density around 
the scatterer, as shown in Fig. \ref{density} for $k_{\mathrm F}=\pi/32$ 
(filling factor $1/32$). Since the Hartree term is exactly compensated 
in our model, those oscillations are only due to the exchange energy.

\subsection{$\langle c_{p+1}^{\dagger}c_p \rangle$ outside the scatterer}

 The conductance $g_1^{k_{\mathrm{F}}} (v)$ was obtained assuming infinite 
perfect leads outside the central scattering region. To know to what extend 
this condition does really matter, we calculate 
$\langle c_{p+1}^{\dagger}c_p\rangle$ outside the two sites where the 
electrons interact. When $p\geq 1$, 
\begin{equation}
\left\langle c_{p+1}^{\dagger}c_p \right\rangle = 
A_{\rm cb}^{p+1,p}+ A^{p+1,p}_{\rm bs1},  
\end{equation}
where the contributions $A^{p+1,p}_{\rm cb}$ of the conduction band 
and $A^{p+1,p}_{\rm bs1}$ of the bound state to 
$\langle c_{p+1}^{\dagger}c_p \rangle$ read:   
\begin{equation}
\begin{aligned}
A^{p+1,p}_{\rm cb}= & \frac{L}{2 \pi} \int_{0}^{k_{\mathrm{F}}} dk  
\left\{\sum_{i=0}^1 \psi_k^{i*} (p+1) \psi_k^i (p)\right\} \\
= & \int_{0}^{k_{\mathrm{F}}} \frac{dk}{\pi} \left\{ \cos k - G(v,k) \right\} 
\\
G(v,k)= & \left(v^2-1 \right) 
\frac{v^2 \cos(2kp-k)-\cos(2kp+k)}{1+v^4-2v^2\cos(2k)} \\ 
A^{p+1,p}_{\rm bs1} = & \frac{v^2-1}{2} v^{-2p-1}. 
\label{cpcp}
\end{aligned}
\end{equation}
\begin{figure}
\centerline{\includegraphics[width=\linewidth]{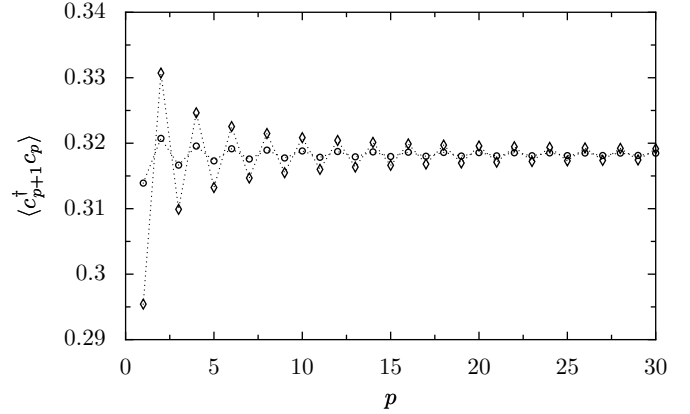}}
\caption{Oscillatory decay of $\langle c_{p+1}^{\dagger}c_p \rangle$ 
towards its asymptotic value $1/\pi$ as a function of $p$ for 
$k_{\mathrm{F}}=\pi/2$, $U=0.1$ (circles) and $U=0.5$ (diamonds)
obtained by using (\ref{outside})-(\ref{hypergeometric}).}
\label{Fig2bis}
\end{figure}

 After integration, one obtains: 
\begin{equation}
\left\langle c_{p+1}^{\dagger}c_p \right\rangle =   
\frac{\sin k_{\mathrm{F}}}{\pi} +\frac{v^2-1}{\pi(2p+1)} X(k_{\mathrm{F}},p,v) 
\label{outside}
\end{equation}
where the function $X(k_{\mathrm{F}},p,v)$ is defined as: 
\begin{equation}
X(k_{\mathrm{F}},p,v) = {\rm Im} F(k_{\mathrm{F}},p,v) 
\label{defX}
\end{equation}
\begin{equation}
F(k_{\mathrm{F}},p,v) = _2F_1(1,\frac{1}{2}+p,\frac{3}{2}+p,
v^2 \mathrm{e}^{2\mathrm{i}k_{\mathrm{F}}}) 
\mathrm{e}^{\mathrm{i}k_{\mathrm{F}}(2p+1)} \nonumber, 
\end{equation}
$_2F_1(\alpha,\beta,\gamma,z)$ being the Gauss hypergeometric function 
\begin{equation}
_2F_1(\alpha,\beta,\gamma,z)=\sum_{n=0}^{\infty} \frac{(\alpha)_n (\beta)_n}
{(\gamma)_n n!} z^n
 \label{hypergeometric}
\end{equation}
with $(x)_n=\prod_{m=0}^{n} (x +m)$. Using an expansion of the 
Gauss hypergeometric function valid in the limit  $p \rightarrow \infty$: 
\begin{equation}
_2F_1(1,\frac{1}{2}+p,\frac{3}{2}+p,v^2 \mathrm{e}^{2\mathrm{i}k_{\mathrm{F}}})
\rightarrow \frac{1}{1-v^2 \mathrm{e}^{2\mathrm{i}k_{\mathrm{F}}}},
\end{equation} 
one obtains for the behavior of $\langle c_{p+1}^{\dagger} c_p \rangle$ 
far from the scatterer a simpler expression: 
\begin{equation}
\left\langle c_{p+1}^{\dagger}c_p \right\rangle \rightarrow 
\frac{1}{\pi}  \left \{ \sin k_{\mathrm{F}} + \frac{H_{k_{\mathrm{F}}}^v(p)}{2p+1}\right\} 
\end{equation}
where $H_{k_{\mathrm{F}}}^v(p)$ is an oscillatory function given by 
\begin{equation}
H_{k_{\mathrm{F}}}^v(p)=\frac{(v^2-1)\left[-v^2 \sin (k_{\mathrm{F}}(2p-1))+\sin(k_{\mathrm{F}}(2p+1))\right]}
{1+v^4-2v^2 \cos(2k_{\mathrm{F}})}. 
\nonumber 
\end{equation}

  Outside the region where the electrons interact, 
$\langle c_{p+1}^{\dagger}c_p \rangle$ exhibits oscillations of 
periodicity $\pi/k_{\mathrm{F}}$ which have a slow power law decay 
towards an asymptotic value $\sin k_{\mathrm{F}}/\pi$. Those oscillations 
are illustrated in Fig. \ref{Fig2bis} for $k_{\mathrm{F}}=\pi/2$ and two 
values of $U$.  

\section{Transmission through two many-body scatterers coupled via 
conduction electrons}
\label{section3}

\begin{figure}
\centerline{\includegraphics[width=\linewidth]{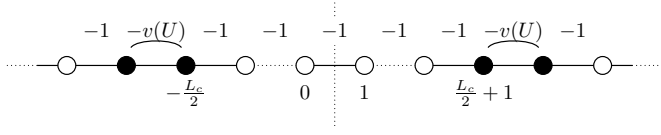}}
\caption{Effective one body model describing two identical many-body 
scatterers (Hamiltonian \eqref{hamiltonian2}) coupled by $L_c$ sites 
where the electrons do not interact.}
\label{Fig2}
\end{figure}

  We now consider an infinite tight binding chain with the same kinetic 
Hamiltonian than defined in Eq. \eqref{hamiltonian}, but with an 
interaction Hamiltonian: 
\begin{equation}
\begin{aligned}
H_{int} =& U (n_{\frac{L_c}{2}+2}-V_{1+})(n_{\frac{L_c}{2}+1}-V_{2+}) \\ 
       + & U (n_{-\frac{L_c}{2}}-V_{2+})(n_{-\frac{L_c}{2}-1}-V_{1+}).
\end{aligned}
\label{hamiltonian2}
\end{equation}
 
 This Hamiltonian describes two nano-systems where the electrons interact, 
as previously studied, which are coupled via an ideal lead of $L_c$ 
sites where the electrons do not interact.  Two positive background 
potentials $V_{1+}$ and $V_{2+}$ are introduced to compensate the Hartree 
terms of  the HF equation. The total Hamiltonian is symmetric under the 
reflection $p-1/2 \rightarrow -p+1/2$. At half-filling, one has particle-hole 
symmetry, $V_{1+}=V_{2+}=1/2$ and the density $n_p=1/2$ is uniform. 
As before, the exchange leads to a modified effective hopping term $v$ 
in the HF-approximation. Because of reflection symmetry, this 
modification must be the same for the two scatterers. The corresponding 
one-body model is sketched in Fig. \ref{Fig2}. However, when there are two 
scatterers in series, the value of $v$ characterizing each scatterer 
becomes different from the value of $v$ obtained in the previous section 
when there is a single scatterer. This is due to the indirect exchange 
interaction which takes place between two scatterers coupled via conduction 
electrons. This indirect exchange interaction gives rise to an effect upon 
the quantum conductance of two nano-systems in series, which vanishes only 
if the length $L_c$ of the coupling wire becomes infinite. We first study 
this effect in the limit when $U \rightarrow 0$, before solving exactly the 
Hartree-Fock equation. We give only the results for $L_c$ even. The extension 
to the case where $L_c$ is odd is straightforward. Moreover, for 
$k_{\mathrm{F}}=\pi/2$ and odd values of $L_c$, the conductance of the two 
scatterers in series $g_2^{\pi/2}=1$, independently of $v$. Because of this, 
we just need $v$ for the even value of $L_c$, at half-filling.

\subsection{Expansion in the weak interaction limit}
 
 For writing the self-consistent equation giving the effective hopping term 
$v$ of one scatterer in series with another, we need to calculate 
the ground state expectation value of $\langle 
c_{\frac{L_c}{2}+2}^{\dagger}c_{\frac{L_c}{2}+1} (v,v) \rangle _2$ 
inside one scatterer when another identical scatterer is located at 
the sites $-\frac{L_c}{2}-1$ and $-\frac{L_c}{2}$. This value depends 
of the two modified hopping terms which characterize each scatterer, 
and which are equal because of reflection symmetry. In the limit of a weak 
interaction strength $U$, $v \rightarrow 1^+$ ($ v > 1$) and one can expand: 
\begin{equation}
\begin{aligned}
&\left\langle c_{\frac{L_c}{2}+2}^{\dagger} c_{\frac{L_c}{2}+1} 
(v,v) \right\rangle_2 = \left\langle c_{\frac{L_c}{2}+2}^{\dagger}
c_{\frac{L_c}{2}+1} (1,1) \right\rangle_2 \\ 
& + \left(v-1\right) \left \{ \frac{\partial}{\partial v} 
C_2(L_c,v) \right \} _{v \rightarrow 1^+} + O\left((v-1)^2\right) \\
\end{aligned}
\end{equation}
where
\begin{equation} 
C_2(L_c,v)= \left\langle c_{\frac{L_c}{2}+2}^{\dagger}c_{\frac{L_c}{2}+1} 
(1,v)\right\rangle_2 + 
\left\langle c_{\frac{L_c}{2}+2}^{\dagger}c_{\frac{L_c}{2}+1}
(v,1)\right\rangle_2. 
\nonumber
\end{equation}
We note that the above expansion involves only terms with a single scatterer, 
for which one can write 
\begin{equation}
\begin{aligned}
\left\langle c_{\frac{L_c}{2}+2}^{\dagger}c_{\frac{L_c}{2}+1} (1,1) 
\right\rangle _2  = & \left\langle c_{1}^{\dagger}c_{0}(1)\right\rangle \\ 
\left\langle c_{\frac{L_c}{2}+2}^{\dagger}c_{\frac{L_c}{2}+1}(v,1)
\right\rangle _2 = & \left\langle c_{1}^{\dagger}c_{0}(v) \right\rangle   \\ 
 \left\langle  c_{\frac{L_c}{2}+2}^{\dagger}c_{\frac{L_c}{2}+1} (1,v)
\right\rangle _2 = & \left\langle c_{L_c+3}^{\dagger} 
c_{L_c+2} (v)\right\rangle.
\end{aligned}
\end{equation}

Using  Eqs. \eqref{inside} and \eqref{outside}, one gets when 
$v \rightarrow 1$ ($v>1$) 
\begin{equation}
\begin{aligned}
& \frac{\partial}{\partial v}\left\langle c_{1}^{\dagger}c_{0}(v) 
\right\rangle \rightarrow \frac{1}{2}-\frac{\sin k_{\mathrm{F}}}{\pi} \\
& \frac{\partial}{\partial v} \left\langle c_{p+1}^{\dagger}c_{p}(v)
\right\rangle \rightarrow \frac{2 X(k_{\mathrm{F}},p,1)}{\pi(2p+1)},
\end{aligned}
\end{equation}
the function $X(k_{\mathrm{F}},p,v)$ being defined in Eq. \eqref{defX}.

 In the weak interaction limit, the effective hopping term $v$ 
characterizing each of the two scatterers in series is given by 
the self-consistent equation:
\begin{equation}
1-v \approx -U \left \{\frac{1}{\pi}+\left(v-1\right) 
\frac{\partial}{\partial v} C_2(L_c,v) \right \}_{v\rightarrow 1^+} 
\end{equation}
where $C_2(L_c,v)= \langle c_{1}^{\dagger}c_{0}(v) \rangle +
\langle c_{L_c+3}^{\dagger}c_{L_c+2}(v) \rangle$. One eventually 
obtains: 
\begin{equation}
v \approx 1 + \left \{ 
\frac{\pi(2-U)}{2U}-\frac{2 X(k_{\mathrm{F}},L_c+2,1)}{2L_c+5}+\sin k_{\mathrm{F}} 
\right \}^{-1}. 
\label{approx}
\end{equation}
when $U \rightarrow 0$. One can see that the indirect exchange 
interaction gives rise to a correction which decays with a power law  
as the length $L_c$ of the coupling wire increases. This is 
when $L_c \rightarrow \infty$ that the scatterers become decoupled, 
and characterized by a value for the effective hopping term which 
coincides with the value given by Eq. \eqref{SCHF-single} 
for a single scatterer in the limit $U \rightarrow 0$:
\begin{equation}
v(L_c=\infty, U\rightarrow 0)\approx 1 + 
\frac{2}{2 \sin k_{\mathrm{F}}-\pi + 2\pi/U}.
\end{equation}
 
Equation \eqref{approx} can be written in a simpler form for a 
half-filled chain. Expressing $X(k_{\mathrm{F}},L_c+2,1)$ for 
$k_{\mathrm{F}}=\pi/2$ one obtains for the effective hopping term 
the equation \eqref{approxbis} given in the introduction. 

\subsection{Exact solution of the Hartree-Fock equation}
 
 Let us solve now the HF-equation for $v$ without assuming that  
$v \rightarrow 1^+$. The effective one-body model described in 
Fig. \ref{Fig2} has even and odd standing-wave solutions 
$\psi_k^0 (p)$ and  $\psi_k^1 (p)$, which can be written inside 
the conduction band as: 
\begin{equation}
\begin{aligned}
\psi_k^0 (p) = & \sqrt{\frac{2}{L}} \cos 
\left(k(p-\frac{1}{2})-\delta_0(k)\right)  \\
\psi_k^1 (p) = & \sqrt{\frac{2}{L}} \sin 
\left(k(p-\frac{1}{2})-\delta_1(k)\right),
\end{aligned}
\end{equation}
at the left side of the two scatterers ($p \leq -\frac{L_c}{2}-1$) and
\begin{equation} 
\begin{aligned}
\psi_k^0 (p) = & \sqrt{\frac{2}{L}} \cos
\left(k(p-\frac{1}{2})+\delta_0(k)\right) \\
\psi_k^1 (p) = & \sqrt{\frac{2}{L}} \sin
\left(k(p-\frac{1}{2})+\delta_1(k)\right)  ,
\end{aligned}
\end{equation}
at the right side of the two scatterers ($p\geq \frac{L_c}{2}+2$). 
Between the two scatterers ($-\frac{L_c}{2} \leq p \leq \frac{L_c}{2}+1$), 
the even and odd standing-wave solutions $\psi_k^0 (p)$ and $\psi_k^1 (p)$ 
read 
\begin{equation} 
\begin{aligned}
\psi_k^0 (p) = & \sqrt{\frac{2}{L}} a_0 \cos
\left(k(p-\frac{1}{2})\right) \\
\psi_k^1 (p) = & \sqrt{\frac{2}{L}} a_1 \sin
\left(k(p-\frac{1}{2})\right),
\end{aligned}
\end{equation}
The expressions for the factors $a_0$  and $a_1$ are 
given in Appendix.
 
 When $ v\neq 1$, the scattering gives rise to two phase shifts 
$\delta_0(k)$ and $\delta_1(k)$, which are given by  
\begin{equation}
\tan \delta_0 (k) = \frac{(v^2-1)\left(\cos (k(L_c+2))+\cos k\right)}
{(v^2-1)\sin (k(L_c+2))+(v^2+1)\sin k}  
\end{equation}
and 
\begin{equation}
\tan \delta_1 (k) = \frac{(1-v^2)\left(-\cos (k(L_c+2))+\cos k\right)} 
{(v^2-1)\sin (k(L_c+2))-(v^2+1)\sin k} 
\end{equation}
respectively. 

 In addition, a value of $v > 1$ can give rise to four bound states 
located around the scatterers with energies outside the conduction band. 
We just write those below the conduction band ($E_{\rm bs}<-2$). There is 
an even bound state of energy $E_{\rm bs0}=-2 \cosh K_0$, $K_0$ being the 
real solution of the equation:       
\begin{equation}
 v^2+ v^2 \exp (K_0(L_c+1)) = 1+\exp (K_0(L_c+3)).
\end{equation}
Its wave function reads
\begin{equation}
\psi_{\rm bs0}(p)=A_0 \exp \left( K_0 (p+\frac{L_c+1}{2}) \right) 
\end{equation}
at the left side of the two scatterers ($p \leq -\frac{L_c}{2}-1$), 
\begin{equation}
\psi_{\rm bs0}(p)=A_0 \exp \left(- K_0 (p-\frac{L_c+3}{2}) \right) 
\end{equation}
at the right side of the two scatterers ($p \geq \frac{L_c}{2}+2$), 
and 
\begin{equation}
\psi_{\rm bs0}(p)=A_0 b_0 \cosh \left( K_0 (p-\frac{1}{2}) \right) 
\end{equation}
between the two scatterers ($ -\frac{L_c}{2} \leq p \leq \frac{L_c}{2}+1$).
The expression for the factor $b_0$ is given in Appendix.

If $L_c$ is large enough, the equation:
\begin{equation}
v^2- v^2 \exp \left(K_1(L_c+1)\right)=  1 - \exp \left(K_1(L_c+3)\right)
\end{equation}
has a real solution for $K_1$. In this case, there is also an 
odd bound state below the conduction band of energy 
$E_{\rm bs1}=-2 \cosh K_1$. Its wave function is given by 
\begin{equation}
\psi_{\rm bs1}(p)= - A_1 \exp \left(K_1 (p+\frac{L_c+1}{2})\right) 
\end{equation}
at the left side of the two scatterers ($p \leq -\frac{L_c}{2}-1$), 
\begin{equation}
\psi_{\rm bs1}(p)=A_1 \exp \left(-K_1(p-\frac{L_c+3}{2}) \right) 
\end{equation}
at the right side of the two scatterers ($p \geq \frac{L_c}{2}+2$), 
and 
\begin{equation}
\psi_{\rm bs1}(p)=A_1b_1 \sinh \left(K_1(p-\frac{1}{2})\right) 
\end{equation}
between the two scatterers ($ -\frac{L_c}{2} \leq p \leq \frac{L_c}{2}+1$).
The expression for the factor $b_1$ is given in Appendix.

\begin{figure}
\centerline{\includegraphics[width=\linewidth]{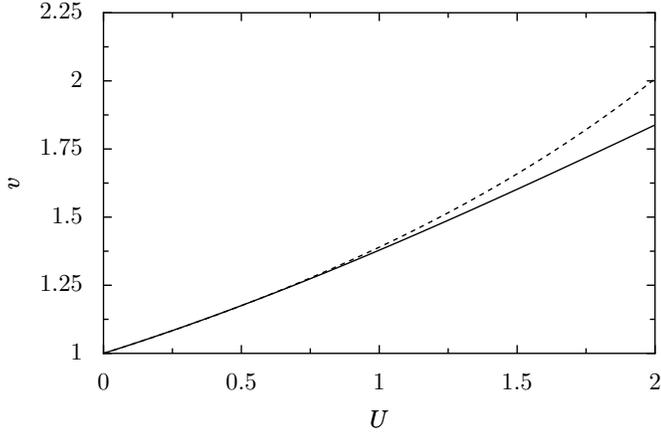}}
\caption{Two scatterers in series: Value $v$ of the effective hopping 
term as a function of $U$. $v$ characterizes  each of the scatterers 
for $L_c=4$ and $k_{\mathrm{F}}=\pi/2$. The solid line gives the exact HF 
value obtained from Eq. \eqref{exact} and the dashed line gives the 
approximated value (Eq. \eqref{approx}).}
\label{Fig3}
\end{figure}
\begin{figure}
\centerline{\includegraphics[width=\linewidth]{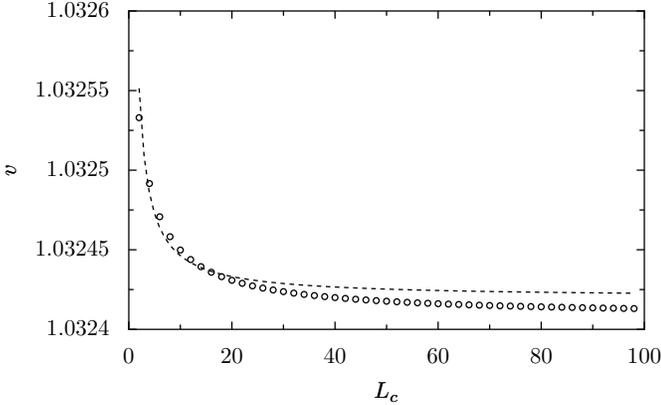}}
\caption{Two scatterers in series: $v$ as a function of even 
values of $L_c$ for $U=0.1$ and $k_{\mathrm{F}}=\pi/2$. 
The circles give the exact HF value obtained from Eq. \eqref{exact} 
and the dashed line gives the approximated value (Eq. \eqref{approx}). 
The even-odd oscillations characteristic of $k_{\mathrm{F}}=\pi/2$ are not 
shown, $v$ for odd $L_c$ being not plotted.}
\label{Fig4}
\end{figure}

The condition 
\begin{equation}
\sum_{p=-\infty}^{\infty} |\psi_{\rm bs0,1}(p)|^2=1
\end{equation}
gives the normalization factors $A_{0,1}$. 
The obtained expressions are somewhat involved and given in Appendix.
 
Using Eq. (\ref{sum-int}), one can calculate
\begin{equation}
\begin{aligned}
\left\langle c_{p+1}^{\dagger} c_{p} \right\rangle 
_2 = & \frac{L}{2\pi} \int_{0}^{k_{\mathrm{F}}} dk 
\left\{ \sum_{i=0}^1 \psi_k^{i*}(p+1) \psi_k^i (p) \right\} \\
& + \sum_{\rm bs} \psi_{\rm bs}^*(p+1) \psi_{\rm bs}(p) 
\end{aligned}
\label{exact}
\end{equation}
for $p=\frac{L_c}{2}+1$   
to obtain the HF equation giving $v$ for two scatterers in series. 
The above integrals have been calculated using Mathematica. 

We compare in Fig. \ref{Fig3} and Fig. \ref{Fig4} the exact HF value 
of $v$ obtained with formula \eqref{exact} to the approximated value 
given by formula \eqref{approx} in the limit $v \rightarrow 1^+$.
One can see that the approximated value is accurate enough for the 
values of $U$ where one can trust the HF approximation. Fig. \ref{Fig4} 
shows the effect of indirect exchange interaction upon the value $v$ of 
the effective hopping, and how this effect disappears when 
$L_c \rightarrow \infty$, for even $L_c$ only.  

\subsection{Density oscillations and quantum conductance for two 
many-body scatterers in series}

\begin{figure}
\centerline{\includegraphics[width=\linewidth]{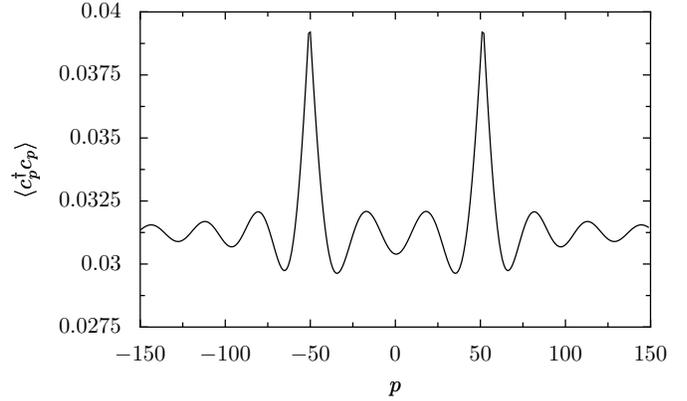}}
\caption{Two scatterers in series: Friedel oscillations of the density  
$n_p=\langle c_{p}^{\dagger}c_p \rangle$ for $U=0.4$, $L_c=100$ and 
$k_{\mathrm{F}}=\pi/32$.}
\label{Density2}
\end{figure}

 Once the self-consistent value for $v$ is obtained, one can calculate 
the Friedel oscillations of the density $n_p=\langle c_{p}^{\dagger}c_p 
\rangle$ around the two scatterers. Outside half-filling, our model 
exhibits Friedel oscillations of the density which are illustrated in 
Fig. \ref{Density2} for $U=0.4$, $L_c=100$ and $k_{\mathrm{F}}=\pi/32$. 
\begin{figure}
\centerline{\includegraphics[width=\linewidth]{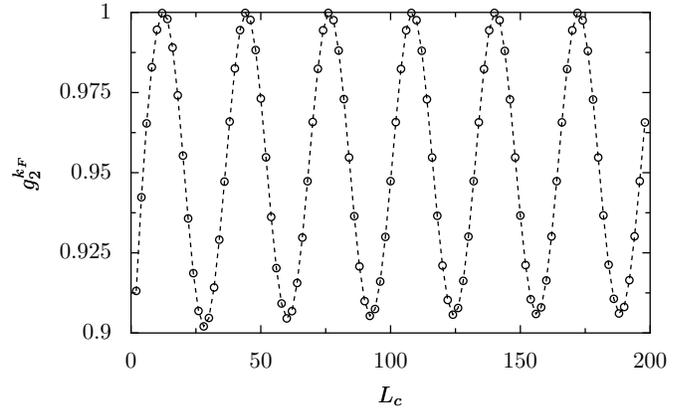}}
\caption{Two scatterers in series: total conductance 
$g_2^{k_{\mathrm{F}}}(L_c,U)$ as a function of $L_c$ 
for $U=0.4$ and $k_{\mathrm{F}}=\pi/32$. Only the values given 
by the exact solution of the Hartree-Fock equation where $L_c$ is even 
are plotted.}
\label{Fig5}
\end{figure}
\begin{figure}
\centerline{\includegraphics[width=\linewidth]{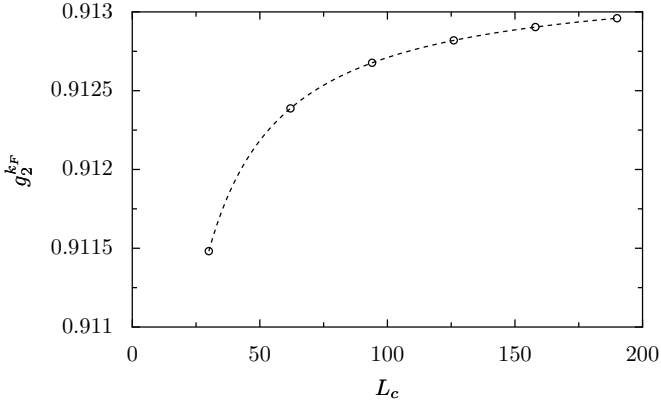}}
\caption{Two scatterers in series: Minimum values of 
$g_2^{k_{\mathrm{F}}}(L_c,U)$ for the successive conductance oscillations 
occurring when $U=0.4$ and  $k_{\mathrm{F}}=\pi/16$, underlining the power 
law decay of their amplitudes as a function of $L_c$. The dashed line is 
a numerical fit $0.91324 - 0.05267/L_c$.}
\label{Fig5bis}
\end{figure}

 The transmission coefficient $t_1^{k_{\mathrm{F}}}$ of a single scatterer 
as a function of $v$ and $k_{\mathrm{F}}$ is given by Eq. \eqref{g1}. 
$r_1^{k_{\mathrm{F}}}$ denoting its reflection coefficient, the transfer 
matrices $M_1^{k_{\mathrm{F}}}$ and $M_{L_c}^{k_{\mathrm{F}}}$ 
through a single scatterer and the coupling lead read: 
\begin{equation}
M_1^{k_{\mathrm{F}}}=\left(\begin{array}{cc} (1/t_1^{k_{\mathrm{F}}})^* 
& r_1^{k_{\mathrm{F}}}/t_1^{k_{\mathrm{F}}} 
\cr (r_1^{k_{\mathrm{F}}}/t_1^{k_{\mathrm{F}}})^* 
& 1/t_1^{k_{\mathrm{F}}}
\end{array}
\right)\, 
\end{equation}
and 
\begin{equation}
M_{L_c}^{k_{\mathrm{F}}}=\left(\begin{array}{cc} 
e^{\mathrm{i}k_\mathrm{F}L_c} & 0 \cr 0 & e^{-\mathrm{i}k_\mathrm{F}L_c} 
\end{array}
\right)\,.
\end{equation} 

 From the matrix $M_2^{k_{\mathrm{F}}}=M_1^{k_{\mathrm{F}}} 
M_{L_c}^{k_{\mathrm{F}}} M_1^{k_{\mathrm{F}}}$, and current conservation  
($1=|t_1^{k_{\mathrm{F}}}|^2 +|r_1^{k_{\mathrm{F}}}|^2$), 
one obtains the transmission coefficient $t_2^{k_{\mathrm{F}}}$ and 
the dimensionless conductance $g_2^{k_{\mathrm{F}}}$ of two scatterers 
in series, coupled by $L_c$ sites where the electrons do not interact: 
\begin{equation}
\begin{aligned}
t_2^{k_{\mathrm{F}}}(v)=& -\frac{2\mathrm{i}v^2 \mathrm{e}^{2\mathrm{i} 
k_{\mathrm{F}}} \sin^2 k_{\mathrm{F}}} {d^{k_{\mathrm{F}}}(v)}
\\
g_2^{k_{\mathrm{F}}}(v)=& \frac{4 v^4\sin^4 k_{\mathrm{F}}}
{\left|d^{k_{\mathrm{F}}}(v)\right|^2},
\label{g2}
\end{aligned}
\end{equation}
where
\begin{equation}
\begin{aligned}
d^{k_{\mathrm{F}}}(v)=&\mathrm{e}^{-\mathrm{i} k_{\mathrm{F}}}
\sin[k_{\mathrm{F}}(L_c+3)] -2v^2 \sin[k_{\mathrm{F}}(L_c+2)]\\
& +v^4 \mathrm{e}^{\mathrm{i} k_{\mathrm{F}}}
\sin[k_{\mathrm{F}}(L_c+1)]. \nonumber
\end{aligned}
\end{equation}

The presence of $L_c$ dependent corrections to $v$ shows that the 
transmission coefficient $t_1^{k_{\mathrm{F}}}$ of one scatterer 
depends on the distance $L_c$ from the other scatterer. Accordingly, 
this non-local effect affects the conductance $g_2^{k_{\mathrm{F}}}$ 
of two scatterers in series. We show in Fig. \ref{Fig5} the total 
conductance $g_2^{k_{\mathrm{F}}}$ of two scatterers in series for a 
low filling factor ($k_{\mathrm{F}}=\pi/32$), obtained by the exact 
solution of the Hartree-Fock equation. One can see the large period 
$\pi/k_{\mathrm{F}}=32$ of the conductance oscillations, though the 
values for odd $L_c$ are not plotted. The conductance oscillations are 
larger when $L_c$ is small, due to the effect  of the exchange energy. 
The $1/L_c$-decay of the conductance oscillations towards its asymptotic 
$L_c$-independent value is underlined in Fig. \ref{Fig5bis} obtained 
with $k_{\mathrm{F}}=\pi/16$. 

\section{Comparison with exact DMRG results}
\label{section4}

Exact values of the conductance of a one dimensional scatterer 
in which electrons interact can be obtained using the embedding method 
and the DMRG algorithm, as explained in previous works 
\cite{MWP,Molina1,Molina2,Molina3}. We have compared in 
Fig. \ref{Fig2} the HF results and the exact DMRG results 
for a single very short scatterer. The difference was negligible for 
small values of $U$. Nevertheless, the HF values differ \cite{Asada} 
more and more from the exact values when the size of the scatterer in which 
the electrons interact increases. The difference should become more 
pronounced for two scatterers in series. We study the ability of the 
HF approximation to describe two interacting nano-systems in 
series for a half filled chain ($k_{\mathrm{F}}=\pi/2$).

If the scattering matrix of one scatterer is not modified by
the presence of other scatterer as for non-interacting systems,
the conductance $g_2^{\pi/2}$ of two scatterers
in series for $k_{\mathrm{F}}=\pi/2$ shows even-odd oscillation 
as a function of the size $L_c$ of the coupling wire: $g_2^{\pi/2} =1$ 
when $L_c$ is odd, and is given by \cite{MWP}
\begin{equation}
g_2^{\pi/2}=\left(\frac{g_1^{\pi/2}}{g_1^{\pi/2}-2}\right)^2 
\label{g2bis}
\end{equation}
when $L_c$ is even. Here $g_1^{\pi/2}$ is the conductance
of a chain with a single scatterer. In the presence of
electron-electron interaction, however, the scattering matrix
of one scatterer can be affected by the other scatterer as shown in the 
previous subsection.

\begin{figure}
\includegraphics[width=\linewidth]{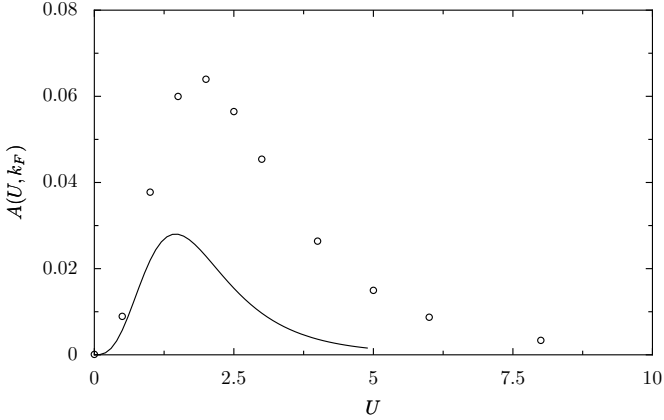}
\caption{Function $A(U,k_{\mathrm{F}})$ characterizing the interaction 
induced correction  $\delta g_2^{k_{\mathrm{F}}} (L_c)$ to 
$g_2^{k_{\mathrm{F}}}$ of (\ref{g2bis}) for two  interacting nano-systems 
in series (formula \eqref{error}) at $k_{\mathrm{F}}=\pi/2$. The exact 
values obtained from the embedding method and the DMRG algorithm  are 
shown by circles, the values obtained from the HF approximation by a 
solid line.} 
\label{Fig6}
\end{figure}

In Ref. \cite{MWP}, the exact values of $g_1^{\pi/2}$ and $g_2^{\pi/2}$ 
were obtained separately, using the embedding method: $g_1^{\pi/2}$ 
being calculated for an infinite chain embedding a single scatterer,   
$g_2^{\pi/2}$ for an infinite chain embedding two scatterers. It was 
found that $g_2^{\pi/2}=1$ if $L_c$ is odd, as predicted. But 
if $L_c$ is even,  $g_2^{\pi/2}$ was related to $g_1^{\pi/2}$ 
by formula \eqref{g2bis} only in the limit $L_c \rightarrow 
\infty$. For small sizes $L_c$, formula \eqref{g2bis}, with 
$g_1^{\pi/2}$ obtained  for a single scatterer surrounded by infinite 
leads without other scatterers, overestimates $g_2^{\pi/2}$ by an amount 
\begin{equation}
\delta g_2^{\pi/2} (L_c) = \frac{A(U,k_{\mathrm{F}})}{L_c} 
\label{error}
\end{equation}
characterized by a function $A(U,k_{\mathrm{F}})$ given in Fig. \ref{Fig6} for 
$k_{\mathrm{F}}=\pi/2$. 
 
 Using the HF approximation, we have shown that the parameter $v$ 
characterizing a single scatterer becomes modified if another scatterer 
is put in series. Hence, the conductance $g_1^{\pi/2}$ of one scatterer 
in series with another differs from its value when it is alone. If 
one ignores this difference, as previously using the DMRG algorithm, 
taking for $g_1^{\pi/2}$ its value without the second scatterer and using 
formula \eqref{g2bis}, one overestimates also $g_2^{\pi/2}$ by an amount 
which is described by formula \eqref{error}, but with a different function 
$A(U,k_{\mathrm{F}})$ given in Fig. \ref{Fig6}. One can see that the
HF approximation reproduces qualitatively the DMRG results, giving
a function $A(U,k_{\mathrm{F}})$ 
characterizing the correction $\delta g_2^{\pi/2} (L_c)$ which first 
increases before decreasing as $U$ varies. When $U$ is small enough, the 
HF approximation reproduces quantitatively the DMRG results: 
$A_{\rm HF}(U,k_{\mathrm{F}}) \approx A_{\rm DMRG}(U,k_{\mathrm{F}})$. 
When $U$ becomes larger, the decay of $A_{\rm DMRG}(U,k_{\mathrm{F}})$ is 
not quantitatively reproduced by the HF approximation. A more suitable 
description could be obtained using a perturbative approach adapted to 
the limit $U \gg t_h$, as used in Ref. \cite{Vasseur}. For intermediate 
$U$, there is no simple analytical approach, making the use of numerical 
renormalization methods (NRG \cite{Hewson,Oguri} or DMRG) necessary. 

\section{Temperature dependent scale $L_\mathrm{T}$ of the 
indirect exchange interactions}
\label{section5}

 We have shown that the indirect exchange interaction between 
two scatterers gives rise to corrections to the value of the 
quantum conductance $g_2^{k_{\mathrm{F}}}$ which slowly decays as $L_c$ 
increases at zero temperature. We are going to show that this effect 
is suppressed when the length of the wire coupling the two scatterers 
exceeds $L_\mathrm{T}$, which is the scale on which an electron 
propagates at the Fermi velocity during a time $\hbar/k_BT$, i.e. 
the thermal length characteristic of free fermions in one dimension. 
Since our approach is essentially valid in the limit of weak values of $U$, 
it is sufficient to consider the weak interaction limit discussed 
in subsection 3.1 for a temperature $T=0$. In this limit, the effect 
was given by the deviation of $\langle c_{p+1}^{\dagger}c_p \rangle$ 
from its uniform value $\sin k_{\mathrm{F}}/\pi$, deviation induced inside 
the leads by an embedded scatterer. To show that this deviation is 
exponentially suppressed above $L_\mathrm{T}$ is enough for proving 
that the effect of the indirect exchange interaction upon 
$g_2^{k_{\mathrm{F}}}$ vanishes when $L_c > L_\mathrm{T}$.   

  At finite temperature $k_BT=\beta^{-1}$, the Fermi-Dirac 
function $f(E,\mu_{\mathrm{F}})$ gives the occupation number of the level of 
energy $E$ at a Fermi chemical potential $\mu_{\mathrm{F}}$:  
\begin{equation}
f_{\beta}(E,\mu_{\mathrm{F}})= 
\frac{1}{\mathrm{e}^{\beta(E -\mu_{\mathrm{F}})}+1}. 
\end{equation}
 When one has a single scatterer embedded in an infinite perfect 
lead, the temperature modifies \cite{Fetter-Walecka} the value of 
$\langle c_{1}^{\dagger}c_{0}\rangle$ inside the scatterer. 
Instead of having Eq. \eqref{inside}, one now has:  
\begin{equation}
\langle c_{1}^{\dagger}c_{0}\rangle_{\beta}= 
A^{1,0}_{\rm cb}(\beta,\mu_{\mathrm{F}})
+A_{\rm bs}^{1,0}(\beta,\mu_{\mathrm{F}})
\end{equation}
where the contribution of the conduction band reads: 
\begin{equation}
\begin{aligned}
A^{1,0}_{\rm cb}(\beta,\mu_{\mathrm{F}}) = & \frac{L}{2\pi}\int_{0}^{\pi} dk 
f_{\beta}(E_k,\mu_{\mathrm{F}})\left\{ \sum_{i=0}^1 \psi_k^{i*} (1) 
\psi_k^i (0) \right\} \\
= & \int_0^{\pi} \frac{dk}{\pi} f_{\beta}(E_k,\mu_{\mathrm{F}})
\left\{ \frac{4v \cos k \sin^2 k}
{1+v^4-2v^2 \cos (2k)} \right\},
%\frac{\cos (3k)-\cos k}{v^4-2v^2 \cos(2k)+1}, 
\nonumber
\end{aligned}
\end{equation}
the contribution of the two bound states becoming 
\begin{equation}
\begin{aligned}
A_{\rm bs}^{1,0}(\mu_{\mathrm{F}}) = & \sum_{\rm bs} f_{\beta}(E_{\rm bs},\mu_{\mathrm{F}}) 
\psi_{\rm bs}^*(1) \psi_{\rm bs}(0) \\
= & \left \{ f_{\beta}(E_{\rm bs1},\mu_{\mathrm{F}}) -f_{\beta}(E_{\rm bs2},\mu_{\mathrm{F}}) 
\right \} \frac{1-v^{-2}}{2}.
\nonumber
\end{aligned}
\end{equation}

\begin{figure}[htb]
\includegraphics[width=\linewidth]{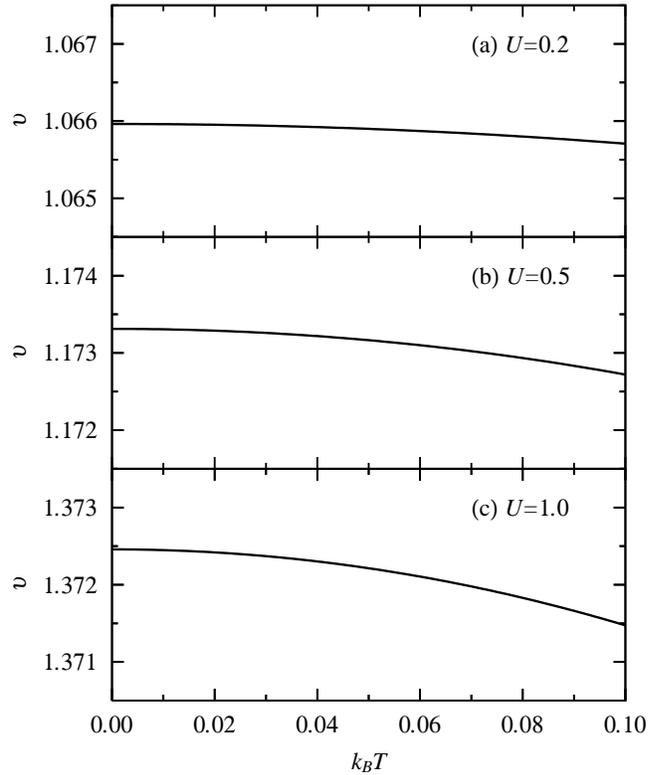}
\caption{$v$ given by Eq. \eqref{eq:scfiniteT} as a function 
of the temperature $T=\beta^{-1}$ for different values of $U$ and 
$\mu_{\mathrm{F}}=0$.}
\label{fig7}
\end{figure}

 The effective hopping term $v$ is given by the implicit equation:
\begin{eqnarray}
v=1+U\langle c_1^{\dagger}c_0 (v)\rangle_{\beta},
\label{eq:scfiniteT} 
\end{eqnarray}
and becomes dependent on the temperature $T$ and on $\mu_{\mathrm{F}}$.

 Assuming a half-filled chain ($\mu_{\mathrm{F}}=0$ due to 
particle-hole symmetry), $v$ given by Eq. \eqref{eq:scfiniteT} has been 
calculated numerically for weak values $U$ of the interaction strength. 
The $T$ dependence of $v$ is shown in Figs.~\ref{fig7}. One can see 
that this dependence remains weak when the temperature $k_BT \leq 0.1$ and 
$U$ is small.

The equation \eqref{outside} giving $\langle c_{p+1}^{\dagger} c_{p} 
\rangle$ outside the part where the electrons interact ($p\geq 1$) 
becomes at a temperature $T$: 
\begin{eqnarray}
\left\langle c_{p+1}^{\dagger} c_{p}\right\rangle_{\beta}
&=& \frac{L}{2\pi} \int_{0}^{\pi} dk f_{\beta}(E_k,\mu_{\mathrm{F}})
\left\{\sum_{i=0}^1 \psi_k^{i*} (p+1)\psi_k^i (p)\right\}  \nonumber 
\\
&& + \sum_{\rm bs} f_{\beta}(E_{\rm bs},\mu_{\mathrm{F}}) 
\psi_{\rm bs}\left(p+1\right) \psi_{\rm bs}\left(p\right)  
\\
&=&\frac{1}{\pi}\int_{0}^{\pi} dk f_{\beta}(E_k,\mu_{\mathrm{F}}) 
\left \{ \cos k - G(k,v) \right \}
\nonumber \\ &&
\nonumber \\&&
+\left \{ f_{\beta}(E_{\rm bs1},\mu_{\mathrm{F}}) - f_{\beta}(E_{\rm bs2},\mu_{\mathrm{F}}) 
\right \} \frac{v^2-1}{2} v^{-2p-1}, \nonumber 
\end{eqnarray}
the function $G(k,v)$ being defined in Eq. \eqref{cpcp}.

\begin{figure}
\includegraphics[width=\linewidth]{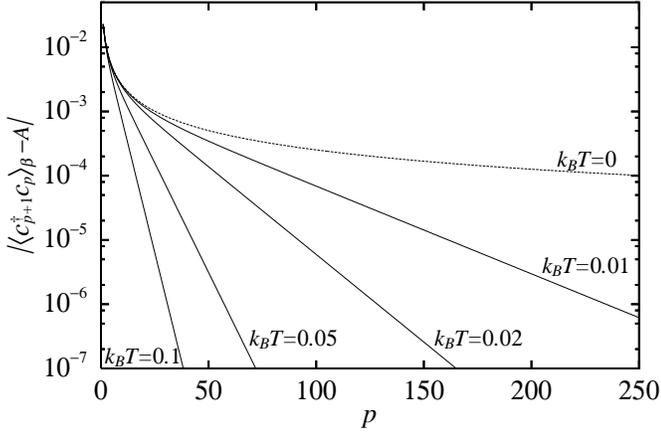}
\caption{
$\left| \left\langle c_{p+1}^{\dagger}
c_{p}\right\rangle-A \right|$ as
a function of $p$ for $U=0.5$ and $\mu_{\mathrm{F}}=0$ (half-filling). 
$A$ is the asymptotic value when $p \rightarrow \infty$.}
\label{Fig8}
\end{figure}

 By using numerical integration, we have calculated $\langle 
c_{p+1}^{\dagger} c_{p} \rangle_{\beta}$ as a function of $p$ for different 
temperatures $T$ and $\mu_{\mathrm{F}}=0$. The data show that $\langle 
c_{p+1}^{\dagger} c_{p} \rangle_{\beta}$ exhibits oscillations which 
have a faster decay when the temperature increases. If $A$ is 
the asymptotic value of $\langle c_{p+1}^{\dagger} c_{p} \rangle_{\beta}$ 
when $p \rightarrow \infty$, we show in Fig . 
\ref{Fig8} how the absolute value of the difference 
between $\langle c_{p+1}^{\dagger} c_{p} \rangle_{\beta}$ and $A$ decays 
when $p$ increases for different values of  $T$.
One can see that the decay towards its asymptotic value of 
$\langle c_{p+1}^{\dagger}c_{p} \rangle_{\beta}$ becomes exponential: 
\begin{eqnarray}
\left | \left\langle c_{p+1}^{\dagger} c_{p}\right\rangle_{\beta} -A \right |
 \propto \exp \left(-\frac{p}{L_\mathrm{T}}\right)
\end{eqnarray}
when $p$ is large.

\begin{figure}
\includegraphics[width=\linewidth]{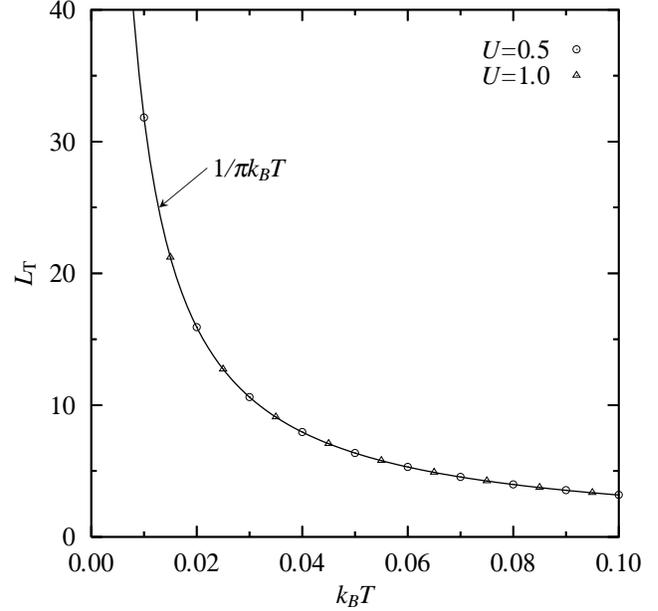}
\caption{Thermal length $L_\mathrm{T}$ as a function of the temperature $T$ 
obtained from the exponential decays of  
$\langle c_{p+1}^{\dagger}c_{p} \rangle_{\beta}$ shown in Fig. \ref{Fig8}.
The points obtained using two values of $U=$ $(0.5$ and $1$) and $T$ form 
a single solid curve $L_\mathrm{T}=1/(\pi k_B T)$.
}
\label{Fig9}
\end{figure}

In Fig.~\ref{Fig9} the decay length  $L_\mathrm{T}$ is shown
as a function of $T$. One can see that $L_T$
decays when the temperature increases as: 
\begin{eqnarray}
L_\mathrm{T}=\frac{1}{\pi k_B T}.
\end{eqnarray}
 Since we have taken $t_\mathrm{h}=1$ and the lattice 
spacing $s=1$ in our 
calculations, one can identify this decay length with the 
length on which a free fermion of speed $v_F=\hbar^{-1} 
\partial E_F/\partial k_{\mathrm{F}} = 2 /\hbar$ propagates during a time 
$\hbar/(k_BT)$ in one dimension. Ignoring a multiplicative factor 
$1/(2\pi)$, $L_\mathrm{T}$ is the usual thermal length of free 
fermions in one dimension.

\section{Conclusion}
\label{section6}

Extending the Landauer formulation of quantum transport to 
nano-systems inside which electrons interact, we have studied a one 
dimensional spinless model which is simple enough to be 
analytically solved assuming the Hartree-Fock approximation. 
We have shown that the scattering  becomes non local when the many-body 
effects inside the scatterer are taken into account. Using two identical 
nano-systems in which interaction gives rise to scattering and which are 
coupled by a non-interacting lead of length $L_c$, the Hartree-Fock 
approximation have allowed us to map the many-body scatterers onto effective 
one-body scatterers which depend on the other scatterer through the 
indirect exchange interaction via the conduction electrons of the coupling 
lead. The non local character of the scattering is only due to exchange 
terms in the studied model, a positive background charge compensating the 
Hartree contribution. We have shown that the HF theory provides 
a qualitative understanding of the non local effects found in Ref.\cite{MWP}, 
quantitatively reproducing the exact behaviors in the weak interaction limit. 

 Eventually, we point out that we have restricted our HF study to a purely 
one dimensional limit where powerful renormalization methods (DMRG, NRG) 
are available. Though it has allowed us to compare the HF results to 
exact numerical results, this is the worst limit for using the HF 
approximation. To extend the HF study to the many channel case, where 
the electron dynamics will be more two or three dimensional does not 
present particular difficulties \cite{Asada}. Moreover, the HF 
approximation is believed to become more accurate outside one dimension. 
A HF study of this many channel limit is in progress and will be 
published in another work.

 We believe that the interaction induced enhancement of the sensitivity 
of the quantum conductance to the nature of the attached leads can be 
relatively easily observed. A more detailed study of a possible experiment 
will be discussed in a separate work.

%%%%%%%%%%%%%%%%Acknowledgments%%%%%%%%%%%%%%%%
\section{Acknowledgments}

One of the authors (Y.A.) is grateful to Research Fellowships of
the Japan Society for the Promotion of Science for Young Scientists.

\section{Appendix}

 For two scatterers in series, one uses even and odd standing 
wave functions.
The factors $a_0$ and $a_1$ for 
the even and odd functions of energy inside the conduction band read:
\begin{equation} 
\begin{aligned}
a_0= & \frac{v |\sin k|}{\sqrt \alpha_0}, \\
\alpha_0= & \left \{ (v^2-1) \cos\left(\frac{k}{2}(L_c+1)\right)
\right \}^2 \\ 
 & + \sin k \left \{ v^2 \sin k + (v^2-1) \sin(k(L_c+2)) \right \}, \\ 
a_1=& \frac{v |\sin k|}{\sqrt \alpha_1}, \\ 
\alpha_1=& \left \{(v^2-1) \sin \left(\frac{k}{2}(L_c+1)\right)
\right \}^2 \\ 
& + \sin k \left \{v^2 \sin k - (v^2-1) \sin (k(L_c+2)) \right \}.
\nonumber
\end{aligned}
\end{equation}
The factors $b_0$ and $b_1$ for the even and odd functions
of energy outside the conduction band read:
\begin{equation} 
\begin{aligned}
b_0 = & \frac{v}{\mathrm{e}^{K_0/2}\cosh  \frac{K_0(L_c+3)}{2}}, \\
b_1 = & \frac{v}{\mathrm{e}^{K_1/2}\sinh  \frac{K_1(L_c+3)}{2}}.
\nonumber
\end{aligned}
\end{equation}

 Using the auxiliary functions $A_{\pm }(L_c,K)$ 
\begin{equation}
A_{\pm }(L_c,K)= 2+L_c \pm \frac {\sinh (k(L_c+2))}{\sinh K}, 
\nonumber
\end{equation}
the normalization factors $A_0$ and $A_1$ for the even and odd bound
states of energy below the conduction band read:
\begin{equation}
\begin{aligned}
\frac{A_0}{\sqrt{2}\mathrm{e}^{\mathrm{i}K_0/2}}= \left \{ 
2+2\coth K_0 + \frac 
{v^2 A_{+}(L_c,K_0)}
{\left(\cosh \frac{K_0(L_c+3)}{2}\right)^2} 
\right \}^{-1/2},
\nonumber
\end{aligned}
\end{equation}
\begin{equation}
\begin{aligned}
\frac{A_1}{\sqrt{2} \mathrm{e}^{\mathrm{i}K_1/2}} = \left \{
2+2 \coth K_1 - \frac
{v^2 A_{-}(L_c,K_1)}
{\left(\sinh \frac{K_1(L_c+3)}{2} \right)^2}
\right \}^{-1/2}.
\nonumber
\end{aligned}
\end{equation}

\end{document}